\documentclass{elsart}
\usepackage{epsfig}
\usepackage{amssymb}

\begin{document}
\begin{frontmatter}
\title{Hard core particle exclusion effects in low dimensional non-equilibrium 
phase transitions}
\author{G\'eza \'Odor$^1$ and N\'ora Menyh\'ard$^2$}
\address{$^1$Research Institute for Technical Physics and Materials Science, 
Budapest, Hungary}
\address{$^2$ Research Institute for Solid State Physics and Optics, 
Budapest, Hungary}
\begin{abstract}
We review the currently known universality classes of continuous 
phase transitions to absorbing states in nonequilibrium systems 
and present results of simulations and arguments to show how the 
blockades introduced by different particle species in one dimension 
cause new robust classes. Results of investigations on the dynamic 
scaling behavior of some bosonic spreading models are reported.

\end{abstract}
\begin{keyword}
nonequilibrium\sep phase transitions\sep particle exclusion
\sep critical behavior
\end{keyword}
\end{frontmatter}

\section{Introduction}\label{intro}

Critical universality is an attractive feature in statistical physics 
because wide range of models can be classified purely in terms of their 
collective  behavior into classes. While in equilibrium systems the factors 
determining these classes are quite well understood, 
in nonequilibrium systems  the situation is less clear. 
The 1+1 dimensional reaction-diffusion systems are important for the 
understanding of the universality classes of non-equilibrium system 
because order-disorder phase transitions may occur in such a low 
dimension unlike in equilibrium systems \cite{DickMar}. 
However, for the ordered state to be stable it must exhibit only small 
fluctuations. All of this kind of transitions are such that the ordered 
state is "absorbing", i.e. when the system falls into it, it can not escape. 
In fact for a long time only such phase transitions have been known 
where the absorbing state is completely frozen. 
A few universality classes of this kind are known \cite{GrasWu},\cite{Hayeof},
the most prominent and the first one that was discovered is that of the 
directed percolation (DP)\cite{DP}.
An early hypothesis \cite{DPuni} was confirmed by all examples up to now.
This claims that in one component systems exhibiting continuous phase 
transitions to single absorbing states (without extra symmetry, inhomogeneity 
or disorder) short ranged interactions can generate DP class transition only.
The Lagragian of the field theory of the DP process $L(\phi(x,t),\psi(x,t))$ 
exhibits a time reversal symmetry $\phi(x,t)\leftrightarrow\psi(x,-t)$.
This results in relations among scaling exponents (see \cite{Hayeof}).
 
The same static scaling behavior was observed in frozen multi-absorbing state
systems as well, like in the pair contact process (PCP) \cite{PCP}. 
Here the time reversal symmetry is broken owing to an extra term in the
Lagrangian describing the effect of frozen particles generating long time 
memory in the system \cite{MGT}.
Besides DP a few more universality classes have been established in the
last decade. In models exhibiting particle annihilation fluctuating 
absorbing states with a single wandering particle can occur \cite{Munoz}.
In Table \ref{tab} we have tried to summarize the most well known absorbing 
state phase transition classes of disorder-free, homogeneous models with 
short range interactions. Those, which are below the horizontal line 
exhibit fluctuating absorbing states. For more details of field theoretical 
symmetries and their relation to hyperscaling laws see \cite{MGT}.
\begin{table}
\caption{Summary of known 1+1 D universality classes. DCF, NDCF denote models 
with coupled (non)diffusive conserved fields.}
\label{tab}
\begin{tabular}{|c|c|c|c|}
\hline
CLASS ID    &  features        & degree of knowledge  &  references\\
\hline\hline
DP          &  time reversal symmetry    & RG to $\epsilon^2$, series exp. & 
\cite{DickMar,GrasWu,Hayeof,DP,DPuni,HayeDPexp} \\
PCP         &  broken time reversal symm. & RG, simulations & \cite{PCP} \\
NDCF        &  global conservation & RG, simulations & 
                                                    \cite{rossi,munoz,pastor}\\
\hline                                              
                                                    
CDP         &  Compact DP, Glauber Ising & exactly solved &  
                                                    \cite{Essam,DoKi,gla63}  \\
PC          &  parity cons.$+$ $Z_2$ symmetry & RG, simulations & 
              \cite{Taka,Jensen,Cardy-Tauber,Gras84,Men94,MeOd96,Park94,Bas,Hin97}\\
N-BARW2     &  N-comp. parity conservation & RG, simu., DMRG &  
                                           \cite{Cardy-Tauber,barw2cikk,Hoy} \\
PCPD        &  binary production   & simulations & 
                   \cite{HT97,Odo00,Carlon,Hayepcpd,Odo01,HayeDP-ARW,binary} \\

DCF         &  global conservation & RG, simulations & 
                                                     \cite{kss,woh,frei,ful} \\

N-BARW2s    & N-comp. sym. BARW + exc.& simulations, MF & 
                            \cite{barw2cikk,Park,dp2cikk,Kor} \\
N-BARW2a    & N-comp. asym. BARW + exc.& simulations, MF &
                            \cite{Park,barw2cikk} \\
\hline          
\end{tabular}
\end{table}

A major problem of these models is that they are usually far from the 
critical dimension and critical fluctuations prohibit mean-field (MF)
like behavior. Further complication is that bosonic field theoretical 
methods can not describe particle exclusion that may obviously happen in 1D. 
The success of the application of bosonic field theory for models shown 
in the first part of the table is the consequence of the 
asymptotically low density of particles near the critical point.
However in multi-component systems, where the exchange between different 
types is non-trivial, bosonic field theoretical descriptions may fail.
Also in case of the binary production (PCPD) models \cite{HT97} it 
predicts a non-vanishing density at the transition point and diverges
in the active phase contrary to the lattice model version of
hard-core particles \cite{Odo00,Carlon,Hayepcpd,Odo01,HayeDP-ARW,binary}. 
Fermionic field theories on the other hand have the disadvantage that 
they are non-local and results exist for very simple reaction-diffusion 
systems only \cite{SPark,Wij}.
Other techniques like independent interval approximation \cite{IIA},
empty interval method \cite{EIm} or density matrix renormalization (DMRG) 
\cite{DMRGH} are currently being developed to be able to solve multi-component
reaction diffusion models. The main aim of this work is to give an overview
and present simulation results for these systems.

\section{Critical dynamical behavior of single-species spreading
processes}

The bosonic field theory of the directed percolation was established by
\cite{GS}. Perturbative $\epsilon=4-d$ renormaliztion group analysis 
\cite{DPuni} up to two-loop order resulted in estimates for the critical 
exponents. For the density decay exponent it gives
\begin{equation}
\alpha_B = 1 - \epsilon/4 - 0.01283 \epsilon^2
\end{equation}
which results in $\alpha_B=0.13453$ in one dimension. This differs from
the most precise simulation and series expansion results of DP 
$\alpha=0.1595(1)$ \cite{IJensen} with about 18\%.
An attempt for fermionic field theoretical solution in 1d (that does not 
allow multiple occupancy of sites) was shown in \cite{Wij} but run into 
severe convergence problems and has not resulted in precise quantitative 
estimates for critical exponents.
Although the bosonic field theory is expected to be valid owing to the 
asymptotically low density at criticality it has never been proven 
rigorously. Contrary, a recent study on a special one-dimensional
DP model predicts different bosonic and fermionic field theoretical 
results \cite{Wij2}. As far as we know all simulations of directed 
percolation (and other spreading models) have been done on lattices 
allowing single occupancy. Moreover there are no more precise estimates 
for the critical exponents of the 1d bosonic model than those of the 
$\epsilon^2$ expansion cited above.
Therefore we decided to perform some simulations with 
bosonic particles to check the scaling behavior at criticality.

\begin{figure}
\epsfxsize=100mm
\centerline{\epsffile{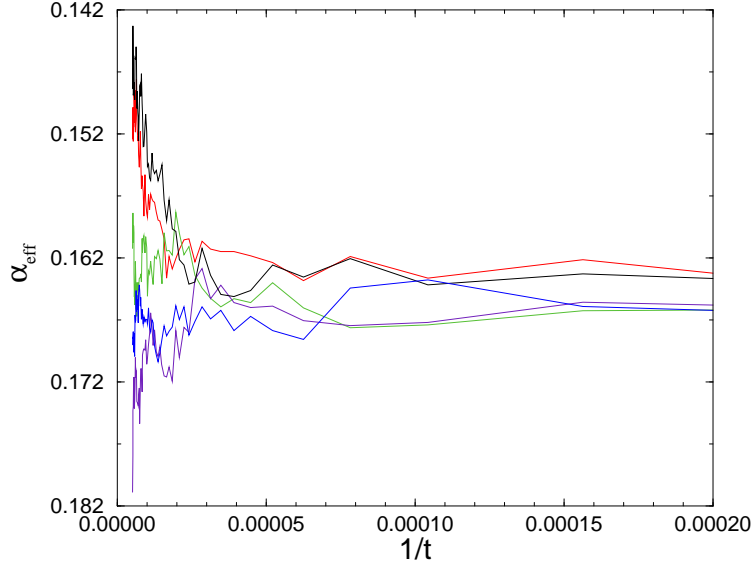}}
\vspace{2mm}
\caption{Local slopes of the density decay in a bosonic DP model.
Different curves correspond to $\lambda=0.12883$, $0.12882$, $0.12881$
$0.1288$, $0.12879$ (from bottom to top).}
\label{bdp}
\end{figure}

First we investigated the following one-dimensional branching and annihilating
process with one offspring (BARW1)
\begin{equation}
A\stackrel{\sigma}{\to} 2A \ \ \ \ \ 2A\stackrel{\lambda}{\to}\emptyset
\ \ \ \ \ A\emptyset\stackrel{1-\sigma-\lambda}{\leftrightarrow}\emptyset A
\end{equation}
in such a way that the branching and the coagulation processes happen
in place. The simulations were run on $L=10^6$ systems with periodic 
boundary conditions and with initially randomly distributed $\emptyset$-s 
and $A$-s of probability 1/2. The density of particles ($\rho(t)$)
is measured up to $t_{max}=2\times 10^5$ Monte Carlo steps (MCS) and 
averaged over $10^4$ samples. By fixing the branching rate to 
$\sigma=0.1$ and varying the
annihilating rate we determined the critical point $\lambda_c$ with the
local slope analysis of data
\begin{equation}
\alpha_{eff}(t) = {- \ln \left[ \rho(t) / \rho(t/m) \right] 
\over \ln(m)} \label{slopes}
\end{equation}
(where we use $m=8$ usually). In the $t\to\infty$ limit the critical curve 
goes to exponent $\alpha$ by a straight line, while in sub(super)-critical 
cases they veer down(up) respectively. As Figure \ref{bdp} shows the critical
point is at $\lambda_c=0.12882(1)$ with $\alpha=0.165(5)$ that agrees well
with other simulation and series expansion results 0.1595(1) \cite{IJensen}.
By simulating on smaller lattice sizes ($L=20000$) we found strong finite 
size effects.

\begin{figure}
\epsfxsize=100mm
\centerline{\epsffile{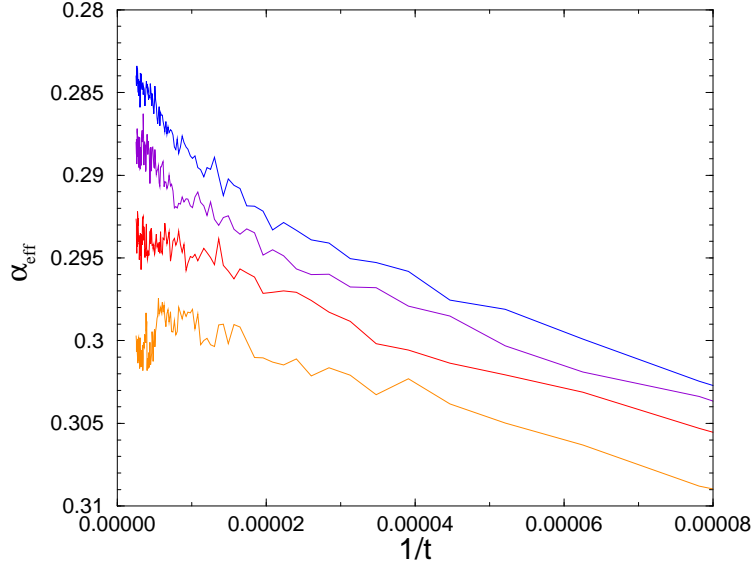}}
\vspace{2mm}
\caption{Local slopes of the density decay in a bosonic BARW2 model.
Different curves correspond to $\sigma=0.466$, $0.468$ $0.469$, 
$0.47$ (from bottom to top).}
\label{bpc}
\end{figure}

Next we performed the same analysis for the 1d two-offspring version BARW
(BARW2)
\begin{equation}
A\stackrel{\sigma}{\to} 3A \ \ \ \ \ 2A\stackrel{\lambda}{\to}\emptyset
\ \ \ \ \ A\emptyset\stackrel{1-\sigma-\lambda}{\leftrightarrow}\emptyset A
\end{equation}
that exhibits mod 2 particle number conservation and its critical behavior
is known to belong to the PC class \cite{Taka,Cardy-Tauber}. Now we fixed
$\lambda=0.2$ and as Fig.\ref{bpc} shows the critical point is at 
$\sigma_c=0.04685(5)$ with the corresponding decay exponent 
$\alpha=0.290(3)$. This value agrees with that of the PC class again 
$0.285(2)$ \cite{Jensen}.

\begin{figure}
\epsfxsize=100mm
\centerline{\epsffile{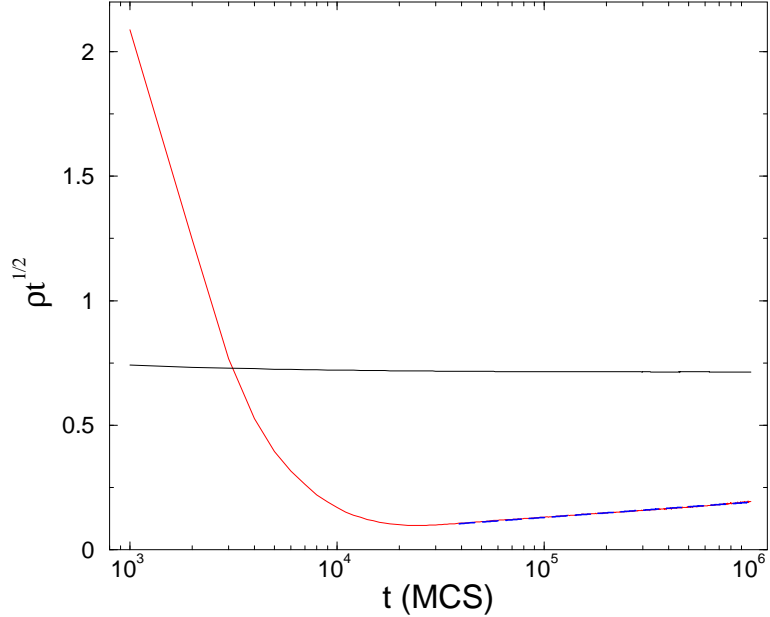}}
\vspace{2mm}
\caption{Density decay in the bosonic annihilation-fission model. The upper
curve corresponds to the inactive phase $\lambda=0.5$, the lower one
to the transition point $\lambda=0.1$. The dashed line shows logarithmic
fitting of the form eq.(\ref{3As}).}
\label{baf}
\end{figure}

Finally we performed bosonic simulations for the sub-critical behavior of 
the following 1d binary spreading process
\begin{equation}
2A\stackrel{\sigma}{\to} 3A \ \ \ \ \ 2A\stackrel{\lambda}{\to}\emptyset
\ \ \ \ \ A\emptyset\stackrel{1-\sigma-\lambda}{\leftrightarrow}\emptyset A
\end{equation}
In the fermionic version of this model continuous phase transition was
found \cite{Odo00,Carlon,Hayepcpd} with new critical behavior.
On the other hand the field theory of the bosonic model \cite{HT97}
predicted discontinuous phase transition at $2\lambda=\sigma$ with diverging 
particle density in the active phase. Our present simulations confirm this. 
While in the inactive phase the $2A\to\emptyset$ process 
(see Section \ref{2Ato0}) governs the evolution it is guessed \cite{Uunp}
that at the transition point the $3A\to\emptyset$ dominates
with  the following scaling law in 1d
\begin{equation}
\rho(t) \propto (\ln(t)/t)^{1/2} + O(1/t) \ \ .
\label{3As}
\end{equation}
The simulations at $\sigma=0.2$ in systems with size $L=10^6$ confirmed 
these expectations  as shown on Fig.\ref{baf}.
At the transition point for $t>4\times10^4$ MCS we applied the form
eq.(\ref{3As}) and found a good fitting with 
$\rho(t) = ((0.0266(1) - 0.1765(1) \ln(t))/t)^{1/2}$.
Though the identification of a universality class requires
the determination of three independent critical exponents we believe 
that  the above numerical data give enough support for 
the expected critical behaviors.

\section{Critical dynamical behavior of multi-species 
annihilating random walks}

First we recall some well known results (see refs. in \cite{Priv})
for simple reaction-diffusion systems, that lack particle creation. 
From the viewpoint of phase transitions this describes the behavior 
in the inactive phase or as we shall show in the N-BARW models right
at the critical point.

\subsection{Models  $A+A\to\emptyset$ and $A+B\to\emptyset$ } \label{2Ato0}

The simplest reaction-diffusion model $A+A\to\emptyset$  (ARW) -- 
in which identical particles 
follow random walk and annihilate on contact
 is exactly solvable in 1D
\cite{Racz85,Lush87}, the particle density decays as
\begin{equation} \label{ARWlaw}
\rho(t)\propto t^{-1/2} \ \ .
\end{equation}
ARW was also shown to be equivalent to the $A+A\to A$ model by 
\cite{Peliti} and renormalization group approach provided universal 
decay amplitudes to all orders in epsilon expansion. 
It was also shown \cite{DoKi} that the motion of kinks in the compact 
version of directed percolation (CDP) \cite{Essam} and in the Glauber-Ising 
model \cite{gla63} at zero temperature  are also described exactly
by (\ref{ARWlaw}).

Two types of particles undergo diffusive random walk and react upon contact
to form an inert particle in the simplest two-component reaction-diffusion 
model $A+B\to\emptyset$.
 For $d<4$ and for  equal initial
density of $A$ and $B$ particles ($\rho_0$), which are randomly situated 
in space the density decays asymptotically as \cite{old}
$\rho_A(t)=\rho_B(t) \propto C_d\sqrt\rho_0  t^{-1/4}$ 
where $C_d$ is a dimension dependent constant. This slow decay  is 
due to the formation of clusters of like particles that do not react and 
their asymptotical segregation for $d<4$. The asymptotically dominant 
process is the diffusive decay of the fluctuations in the initial conditions. 
Since this is a short ranged process the system will have a long-time memory 
-- appearing in the amplitude dependence -- for the initial density.

\subsection{The effect of exclusion} \label{2}

A simple reaction-diffusion model of two types
$A+A\to\emptyset$, $B+B\to\emptyset$ with exclusion 
$AB\not\leftrightarrow BA$ is offered by the
 Generalized Domany-Kinzel (GDK) stochastic cellular automaton model 
of Hinrichsen  \cite{Hin97} introduced originally to realize phase 
transitions from active ($Ac$) to multiple inactive absorbing states: 
$I1$, $I2$. 
We investigated numerically \cite{gdkcikk} this model in a special point 
of its phase diagram where compact domains of $I1$ and $I2$ 
grow separated 
by $Ac I1 = A$ and $Ac I2 = B$ kinks that can not penetrate each other 
(CDP2). There are  special pairwise initial conditions in the model
(because the domains are bounded by kinks of the same type):
     ....A...A...B.B..B.....B..A..A..

The dynamical behavior of critical systems is usually investigated from
two extreme initial conditions: (a) homogeneous system with randomly 
distributed species (b) empty system with a single initial seed of particles.
In the former case (a) and pairwise initial conditions our numerical 
simulations show a density decay of kinks (or particles of the corresponding 
reaction-diffusion system) $\rho\propto t^{-\alpha}$ characterized by a 
power-law with an exponent larger than $\alpha=0.5$ that would have been 
expected in case of two copies of ARW systems that do not exclude each other. 
Furthermore the deviation of $\alpha$ from 0.5 showed an initial density 
dependence. 
We provided a possible explanation based on symmetry between types that a 
marginal perturbation emerges here that causes this non-universal scaling. 
We showed an analogy with the DP confined by parabolic boundary conditions 
\cite{Turban} by assuming that the kinks  exert a parabolic space-time
confinement on the decaying domains. Non-universal scaling can also be 
observed at surface critical phenomena similarly to here where kinks 
produce 'multi surfaces' in the bulk. However simulations and 
independent interval approximations on a similar model predict a 
$t^{-1/2}/\ln (t)$ behavior \cite{Satya}.
Perhaps a fermionic field theoretical study could help to understand better 
the situation. 
In case of random distribution of $A$-s and $B$-s the density decays
slower owing to the accumulation of $AB$ pairs. An exact mapping onto 
the $A+B\to\emptyset$ model predicts $\rho_A=\rho_B\propto t^{-1/4}$
that was confirmed by simulations \cite{barw2cikk}.

In case (b) above one usually measures the survival probability of clusters
originated from a  seed. In critical systems this scales like
$P(t)\propto t^{-\delta}$ .
When we inserted a seed of $I2$-s in the see of $I1$ and $Ac$-s of the GDK 
model at the CDP2 point we found very strong dependence of 
$\delta$ on $\rho_0(I1)$ \cite{gdkcikk}. 
Also the above picture with parabolic boundary conditions has gained further
support here.

\section{Phase transition generated by offspring production}

If we add particle creation by branching to the systems mentioned 
before we may expect a phase transition, where a steady state with 
constant particle density occurs. For $N$ component parity conserving
systems (N-BARW2, $N=2$ types, two offsprings) it was shown by field 
theory \cite{Cardy-Tauber} that the $A\to 3A$ like processes are not 
relevant and the models with $A\to A2B$ and $B\to B2A$ 
(ordering of offsprings is not relevant) branching terms do exhibit 
continuous phase transitions at $\sigma=0$ branching rate. 
The universality class is expected to be independent from $N$ 
and to coincide with that of the $N\to\infty$ (N-BARW2) model.
The critical dimension is $d_c=2$ and for $d=1$ the exponents are
\begin{equation} \label{NBARWe}
\beta=1, \ \ \ Z=2, \ \ \ \alpha=1/2, \ \ \ \nu_{||}=2, \ \ \ \nu_{\perp}=1
\end{equation}
exactly.

\subsection{The parity conserving 2-BARW2 model with exclusion}

The effect of exclusion in the 2-BARW2 model was investigated in 
\cite{barw2cikk} and for $d=2$ the field theoretical predictions 
were confirmed. In one dimension however two phase transitions of
different types were observed depending on the arrangement of 
offsprings relative to the parent. 
Namely if the parent separates the offsprings: 
$A\stackrel{\sigma}{\to}BAB$ (2-BARW2s) the steady state density will be 
higher than in the case when they are created on the same site: 
$A\stackrel{\sigma}{\to}ABB$ (2-BARW2a) for a given branching rate because 
in the former case they are unable to annihilate with each other.
This results in different order parameter exponents for the symmetric 
and the asymmetric cases ($\beta_s=1/2$ and $\beta_a=2$).
This is in contrast to the widespread beliefs that bosonic field theory
(where $AB\leftrightarrow BA$ is possible) can well describe these systems
because in that case \cite{Cardy-Tauber} the scaling behavior is 
different (\ref{NBARWe}).
This finding led \cite{Park} to the conjecture that in one-dimensional
reaction-diffusion systems a series of new universality classes should
appear if particle exclusion is present. 
Note however that only the off-critical exponents are different, since
in both cases the transition is at $\sigma=0$ and the on-critical ones
are those described in Sect.(\ref{2}). 
In \cite{barw2cikk} a set of critical exponents satisfying scaling
relations have been determined for this two new classes shown in 
Table \ref{tab2}.
\begin{table}
\begin{center}
\begin{tabular}{|l|r|r|r|r|}
\hline
process & $\nu_{||}$      & $Z$ & $\alpha$       & $\beta$ \\ 
\hline
N-BARW2 & 2               &  2  & 1/2            & 1       \\
\hline
N-BARW2s &2.0(1)$|$0.915(2)&4.0(2)$|$1.82(2)*&0.25(1)$|$0.55(1)*&0.50(1)\\
\hline
N-BARW2a &8.0(4)$|$3.66(2) &4.0(2)$|$1.82(2)*&0.25(1)$|$0.55(1)*&2.05(10)\\
\hline 
\end{tabular}
\end{center}
\caption{Summary of critical exponents in one dimension in N-BARW like
models. The non-blocking data are quoted from \cite{Cardy-Tauber}. 
Data divided by "$|$" correspond to random vs. pairwise
initial condition cases \cite{gdkcikk}.
Exponents denoted by * exhibit slight initial density dependence.
}
\label{tab2}
\end{table}

\subsection{The parity non-conserving model with exclusion (2-BARW1)}

Hard-core interactions in the two-component, one-offspring production 
model (2-BARW1) were investigated in \cite{dp2cikk}. Without 
interaction between different species one would expect DP class transition. 
By introducing an $AB\not\leftrightarrow BA$ blocking in the
\begin{eqnarray}
A\stackrel{\sigma}{\longrightarrow} AA \ \ \ \
B\stackrel{\sigma}{\longrightarrow} BB \\
AA\stackrel{1-\sigma}{\longrightarrow}\emptyset \ \ \ \ 
BB\stackrel{1-\sigma}{\longrightarrow}\emptyset 
\end{eqnarray}
model a DP class transition at $\sigma=0.81107$ was been found indeed.
On the other hand if we couple the two sub-systems by the production
\begin{eqnarray}
A\stackrel{\sigma/2}{\longrightarrow} AB \ \ \ \ 
A\stackrel{\sigma/2}{\longrightarrow} BA \\
B\stackrel{\sigma/2}{\longrightarrow} AB \ \ \ \ 
B\stackrel{\sigma/2}{\longrightarrow} BA \\
AA\stackrel{1-\sigma}{\longrightarrow}\emptyset \ \ \ \ 
BB\stackrel{1-\sigma}{\longrightarrow}\emptyset 
\end{eqnarray}
without exclusion bosonic field theory \cite{Janssen-col} predicts
unidirectionally coupled DP class \cite{uni-DP} transitions.
With hard-core exclusion a continuous phase transition will emerge 
at rate $\sigma=0$  -- therefore the on-critical exponents will be
the same as described in Sect. (\ref{2}) -- and the order parameter 
exponent was found to be $\beta=1/2$. This assures that this transition 
belongs to the same class as that of the 2-BARW2s model. 
The parity conservation law that is relevant in case of one component 
systems (PC versus DP class) is irrelevant here. 
This finding reduces the expectations for a whole new series of 
universality classes in 1D systems with exclusions. In fact the 
blockades introduced by exclusions generate robust classes. 
In \cite{dp2cikk} a hypothesis was set up that in coupled branching and 
annihilating random walk systems of 
N-types of excluding particles for continuous transitions at $\sigma=0$ 
two universality classes exist, those of 2-BARW2s and 2-BARW2a, 
depending on whether the reactants can immediately annihilate 
(i.e. when similar particles are not separated by other type(s) 
of particle(s)) or not. Recent investigations in similar
models \cite{Kor,Lip} are in agreement with this hypothesis and
the extension to binary spreading processes was proposed in \cite{parwcikk}.

\section{Summary}

We have shown how blockades generated by particle exclusion in some 
one-dimensional reaction-diffusion type systems affect critical scaling 
behavior.

We confirmed by bosonic simulations that the single component unary
BARW1 and BARW2 processes exhibit the same dynamical scaling as the
corresponding models with exclusion. In case of the one component 
binary spreading process the bosonic and fermionic versions differ. 
While the former one has a discontinuous phase transition the latter 
exhibits a continuous transition. We determined numerically the 
dynamic scaling behavior of the bosonic version.

Without particle production initial condition dependent dynamical
scaling was found in multicomponent ARW models. This critical behavior 
was shown to give the set of on-critical exponents if we generate phase
transition by particle branching since the transitions occur at
the $\sigma=0$ rate. The off-critical exponents were found to be
insensitive to parity conservation law and depend only on the 
spatial arrangement of parent and offspring. Two generic universality
classes: 2-BARW2a and 2-BARW2s were explored by numerical simulations.
In these models the hard-core exclusions generate extra fluctuations
in the absorbing state and the passive steady states posses
long-range correlations.

{\bf Acknowledgements:}\\
The authors thank H. Chat\'e, P. Grassberger and U. C. T\"auber for 
their comments.
Support from Hungarian research funds OTKA (Grant Nos. T25286, T023791, 
T034784) Bolyai (Grant No. BO/00142/99) and IKTA (Project No. 00111/2000) 
is acknowledged. 
The simulations were performed on the parallel cluster of SZTAKI and on the
supercomputer of NIIF Hungary.

\end{document}